\documentclass{elsart3}

\usepackage{graphicx}

\usepackage{amssymb}
\begin{document}

\begin{frontmatter}



\title{Magnetic Order and Dynamics in Stripe-Ordered
La$_{2-x}$Sr$_{x}$NiO$_4$}


\author[Oxford]{A.T. Boothroyd\corauthref{cor1}},
\author[Oxford]{P.G. Freeman},
\author[Oxford]{D. Prabhakaran},
\author[ILL]{M. Enderle},
\author[ILL]{J. Kulda}
\corauth[cor1]{Corresponding author. Tel.: +44 1865 272376; fax:
+44 1865 272400.; e-mail: a.boothroyd@physics.ox.ac.uk}

\address[Oxford]{Department of Physics, Oxford University, Oxford, OX1 3PU, United
Kingdom }
\address[ILL]{Institut Laue--Langevin, BP 156, 38042 Grenoble Cedex 9, France}

\begin{abstract}
We have studied magnetic correlations in several compositions of
stripe-ordered La$_{2-x}$Sr$_{x}$NiO$_4$. In this paper we show
how polarized-neutron scattering has helped uncover important
features of the magnetic ordering and spin dynamics. In
particular, polarization analysis has enabled us (1) to
characterize a spin reorientation transition, (2) to identify
anisotropy gaps in the spin excitation spectrum, and (3) to
investigate an anomalous dip in the spin-wave intensity suggestive
of coupling between collective spin and charge excitations.
\end{abstract}

\end{frontmatter}

\section{Introduction}
\label{sec:intro} The layered nickelate family
La$_{2-x}$Sr$_{x}$NiO$_4$ has been the focus of many recent
investigations on account of its stripe-ordered phase. This phase
is characterized by the segregation of charge carriers into an
ordered array of parallel lines separating antiferromagnetic
domains of Ni$^{2+}$ ($S=1$) spins \cite{stripes-nickelates-expt}.
Stripes represent a novel type of complex order, and their
properties have been energetically researched since the discovery
of stripe order in members of the copper-oxide high-$T_{\rm c}$
superconductors \cite{Tranquada-Nature-1995}. Experiments to
establish the basic physics of the stripe phase in model
non-superconducting systems can help determine whether or not
stripes play an important role in the mechanism of
superconductivity in the cuprates.

The nature of the spin and charge order in the stripe phase in
La$_{2-x}$Sr$_{x}$NiO$_4$ has been characterized in detail over
the doping range $0.15\leq x \leq 0.5$ by electron, x-ray and
neutron diffraction \cite{stripes-nickelates-expt}, and it is now
of interest to explore the collective spin and charge dynamics
over the same range of composition. From such studies one can
learn about the nature of the excited states, and hence gain
insight into the microscopic interactions that stabilize stripe
order.

In this paper we describe neutron scattering studies of the
magnetic order and dynamics in a number of
La$_{2-x}$Sr$_{x}$NiO$_4$ compounds. Some of the data for the
$x=0.33$ and $x=0.5$ systems have been reported previously
\cite{Boothroyd-PRB-2003,Freeman-PRB-2002}. Here we focus on
results obtained with polarized neutrons, and emphasize how
certain characteristic features of the data vary with composition.

\section{Experimental details}
\label{sec:expt} All the data presented here were collected on the
IN20 triple-axis spectrometer at the Institut Laue-Langevin. Since
its upgrade in 2001, the flux of polarized neutrons on IN20 has
increased by an order of magnitude \cite{kulda-ApplPhysA-2002}.
This gain has allowed experimentalists to be more ambitious,
especially with regard to inelastic scattering measurements.

IN20 is equipped with focussing arrays of heusler crystals for
both monochromator and analyser. Typical experiments require
measurement of the spin-flip (SF) and non-spin-flip (NSF)
scattering, with the neutron polarization $\bf P$ in the
horizontal scattering plane either parallel or perpendicular to
the scattering vector $\bf Q$, or with $\bf P$ vertical. For the
present measurements the analyser was set to reflect neutrons of
fixed final energy 14.7\,meV.

Single crystals of La$_{2-x}$Sr$_{x}$NiO$_4$ were prepared in
Oxford by the floating-zone method \cite{Prabhakaran-JCG-2002}.
The crystals were all rod-like with masses of 10--15 g.

\section{Results}
\label{sec:results}The first studies we describe concern the
direction of the ordered magnetic moment of the Ni$^{2+}$ spins in
La$_{2-x}$Sr$_{x}$NiO$_4$. The magnetic ordering temperature
$T_{\rm SO}$ of the Ni$^{2+}$ spins in the stripe phase depends on
$x$. For the compositions reported here, $T_{\rm SO} \approx
200$\,K for $x = 0.275$--0.37, and $T_{\rm SO} \approx 80$\,K for
$x = 0.5$. The diagonal stripe pattern is twinned, with stripes
either parallel to the $[1,-1,0]$ or $[1,1,0]$ directions of the
tetragonal lattice (cell parameter $a=3.8\,{\rm \AA}$). These are
described by two-dimensional magnetic ordering wave vectors ${\bf
Q}_{\rm m} = (\frac{1}{2},\frac{1}{2}) \pm (\delta,\delta)$ and
$(\frac{1}{2},\frac{1}{2}) \pm (\delta,-\delta)$, respectively,
where $\delta \approx x/2$ in reciprocal lattice units. In our
samples we observed an equal population of the twin domains.

Using polarized-neutron diffraction, Lee {\it et al}
\cite{Lee-PRB-2001} deduced that the ordered moments at low
temperature are constrained to lie in the NiO layers at an angle
of $27^\circ$ ($x=0.275$) and $53^\circ$ ($x=0.33$) to the stripe
direction. In the case of $x=0.33$, they also found that a
transition takes place on warming through $T_{\rm SR} \sim 50$\,K
such that the spins rotate in the plane towards the stripe
direction through an angle of $\sim 13^\circ$.

\begin{figure}
\includegraphics{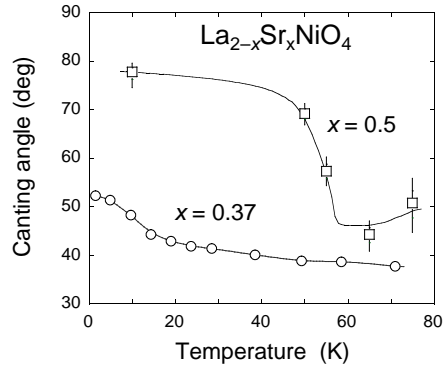}
\caption{Temperature dependence of the in-plane ordered moment
direction in La$_{2-x}$Sr$_{x}$NiO$_4$, $x=0.37$ and 0.5. The
canting angle is defined with respect to the stripe direction.
Lines are visual guides. }
\label{PNSXM_fig1}       
\end{figure}

We have performed similar measurements to Lee {\it et al}, but on
crystals with $x=0.37$ and $x=0.5$. The results are presented in
Fig.~\ref{PNSXM_fig1}, which shows the angle of the spins from the
stripe direction as a function of temperature. Like Lee {\it et
al}, we are assuming in this analysis a collinear magnetic
structure with a single canting angle, as opposed to a
distribution of domains with different spin directions, the
relative population of which could vary. This assumption needs to
be checked out experimentally.

From Fig. ~\ref{PNSXM_fig1}, both $x=0.37$ and $x=0.5$ compounds
are seen to undergo a spin reorientation transition within the NiO
planes, with $T_{\rm SR} \sim 10\,{\rm K}$ $(x=0.37)$ and
$57\,{\rm K}$ $(x=0.5)$. The transition is rather broad for
$x=0.37$. The angle through which the spins rotate is $\sim
13^\circ$ $(x=0.37)$ and $\sim 26^\circ$ $(x=0.5)$

\begin{figure}
\includegraphics{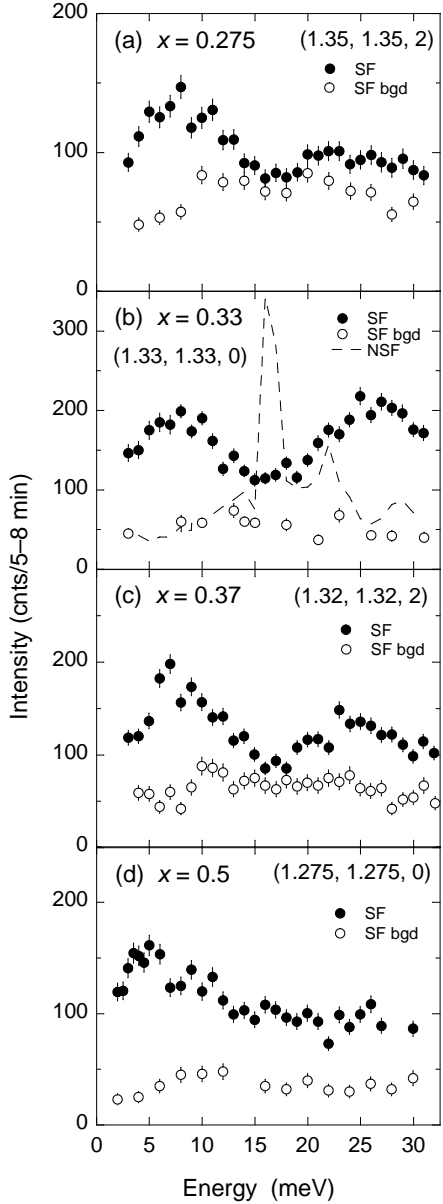}
\caption{Polarized-neutron scattering from stripe-ordered
La$_{2-x}$Sr$_{x}$NiO$_4$. The plots show energy scans performed
at the magnetic ordering wavevector appropriate to each
composition. The spin-flip (SF) scattering was measured with ${\bf
P} \parallel {\bf Q}$, and the SF background was estimated from
scans centred away from the magnetic excitation peak. The
measurements were made at temperatures of 2\,K ($x=0.275, 0.37$),
10\,K($x=0.5$) and 13.5\,K($x=0.33$). }
\label{PNSXM_fig2}       
\end{figure}

Next we discuss two characteristic features observed in the
stripe-phase spin dynamics of La$_{2-x}$Sr$_{x}$NiO$_4$. We refer
to the energy scans shown in Fig.~\ref{PNSXM_fig2}. These scans
were obtained with $\bf Q$ fixed at the two-dimensional magnetic
ordering wavevector ${\bf Q}_{\rm m}$ for each composition. As
expected, cooperative spin excitations emerge from the ${\bf
Q}_{\rm m}$ wavevectors. These excitations are found to be
strongly dispersive, so much so that magnetic scattering is
observed at ${\bf Q}_{\rm m}$ over the whole energy range shown in
Fig.~\ref{PNSXM_fig2}. Inter-plane correlations were observed at
low energies, but were found to be negligible for $E\gtrsim
5$\,meV.

One feature common to all the scans shown in Fig.~\ref{PNSXM_fig2}
is a reduction in intensity below an energy of $\sim 7$\,meV for
$x=0.275-0.37$, or $\sim 4$\,meV for $x=0.5$. This reduction is
not present in energy scans made at equivalent ${\bf Q}_{\rm m}$
wavevectors with a large component along the $c$ axis (not shown).
Given that neutrons scatter from spin fluctuations perpendicular
to ${\bf Q}$, these observations indicate that the intensity
reduction below 4--7\,meV is due to the freezing out of the c-axis
component of the spin fluctuations. This points to the existence
of a 4--7\,meV energy gap due to single-ion out-of-plane
anisotropy.

Polarized neutrons provide a direct method to check for anisotropy
gaps. Below an anisotropy gap one component of the spin
fluctuations is frozen out, and the component involved can be
identified by polarization analysis. We measured the SF scattering
from the $x=0.33$ crystal at energies of 3, 10, 15 and 26\,meV,
with the three-dimensional wavevector fixed at ${\bf Q}_{\rm m} =
(1.33, 1.33, l)$, $l$ = 0 or 1. For these measurements the crystal
was mounted with the $[1, 1, 0]$ and $[0, 0, 1]$ directions in the
horizontal scattering plane, and we switched the neutron
polarization between three orthogonal directions: (i) ${\bf
P}\parallel {\bf Q}_{\rm m}$, (ii) ${\bf P} \perp {\bf Q}_{\rm m}$
in the scattering plane, and (iii) ${\bf P}$ vertical (i.e.\
parallel to the stripe direction $[1, -1, 0]$). We label the
corresponding SF intensities $I_1$, $I_2$ and $I_3$ respectively.
For the case $l=0$, which was used with energies of 10, 15 and
26\,meV, it can be shown that
\begin{equation}
\frac{I_1 - I_2}{I_1 - I_3} = \frac{S_c}{S_\parallel},\label{eq1}
\end{equation}
where $S_c$ and $S_\parallel$ are the components of the dynamic
magnetic response function describing spin correlations along the
crystal $c$ axis ($S_c$), and parallel to the stripe direction
($S_\parallel$). At 3\,meV the intensity at $(1.33, 1.33, l)$
shows a slight modulation with $l$ due to weak inter-layer
correlations. This modulation peaks at odd $l$ integers, and we
made the 3\,meV measurement at the $l=1$ intensity maximum. As far
as the crystal orientation is concerned the difference in angle
between $(1.33, 1.33, 0)$ and $(1.33, 1.33, 1)$ is only $9^\circ$,
and Eq. (\ref{eq1}) is still a very good approximation.

\begin{figure}
\includegraphics{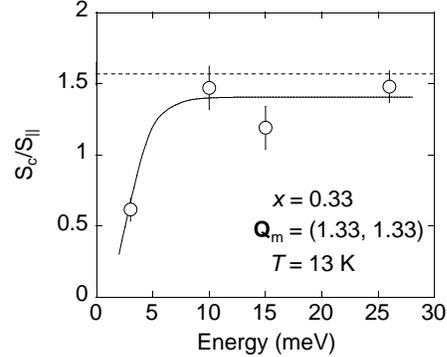}
\caption{Ratio of the magnetic response along the crystal $c$ axis
($S_c$) to that parallel to the stripe direction ($S_\parallel$).
The measurements were made by neutron polarization analysis at the
two-dimensional magnetic zone centre ${\bf Q}_{\rm m} = (1.33,
1.33)$. The broken line is the calculated ratio for isotropic spin
wave oscillations about an ordered moment direction $53^\circ$
away from the stripe direction. The reduction in $S_c/S_\parallel$
at low energies is due to the suppression of out-of-plane spin
fluctuations below the 7\,meV gap visible in
Fig.~\ref{PNSXM_fig2}(b). The solid line is a guide to the eye.}
\label{PNSXM_fig3}       
\end{figure}

The results of these measurements are shown in
Fig.~\ref{PNSXM_fig3}. The ratio $S_c/S_\parallel$ is seen to be
constant within experimental error for energies from 10\,meV to
26\,meV. The average value of $S_c/S_\parallel$ in this energy
range is close to that calculated on the assumption of isotropic
spin fluctuations about the ordered moment direction $53^\circ$
away from the stripe direction \cite{Lee-PRB-2001}. The point at
3\,meV, however, is much lower than the others, indicating a
strong reduction in out-of-plane spin fluctuations at this energy.

These results show that the spin fluctuations are predominantly
in-plane at energies of 3\,meV, and isotropic about the direction
of the ordered moment at 10\,meV and above. This confirms that the
7\,meV feature in Fig.\ \ref{PNSXM_fig2}(b) for the $x=0.33$
crystal is the out-of-plane anisotropy gap. Similar measurements
on the $x=0.5$ crystal showed likewise for the 4\,meV gap in Fig.\
\ref{PNSXM_fig2}(d). We searched for the in-plane anisotropy gap
in energy scans at wavevectors nearly perpendicular to the NiO
layers, but could only put an upper limit on the gap energy of
1\,meV.

Interestingly, the out-of-plane energy gap in undoped
La$_2$NiO$_4$ is 16\,meV \cite{Nakajima-JPSJ-1993}. Together with
the data reported here this indicates that the out-of-plane
anisotropy decreases with doping. In a spin-wave model, the gap
energy varies as $\sqrt{K_c J_{\rm tot}}$, where $K_c$ is the
anisotropy energy and $J_{\rm tot}$ is the total exchange energy
per spin. Measurements of the overall spin-wave dispersion have
established that the intra and inter-stripe Ni--Ni exchange
interactions differ by only a factor 2 \cite{Boothroyd-PRB-2003},
so that $J_{\rm tot}$ decreases relatively slowly with doping.
This implies a very considerable decrease in $K_c$ with doping,
which has been confirmed by spin-wave analyses for $x=0$ and 0.33,
for which $K_c=0.52$\,meV and $0.07$\,meV respectively
\cite{Nakajima-JPSJ-1993,Boothroyd-PRB-2003}.

The other notable feature is the dip centred on $16$\,meV and
subsequent peak near $\sim 26$\,meV in
Figs.~\ref{PNSXM_fig2}(a)--(c). This dip--peak structure occurs in
an energy range where there is strong scattering from phonons, as
evident in the NSF channel shown in Fig.~\ref{PNSXM_fig2}(b), and
so neutron polarization analysis is an important tool to isolate
the magnetic scattering.

Constant-energy scans performed at energies close to 26\,meV
showed that the magnetic response is significantly broader in $\bf
Q$ for $x=0.275$ than for $x=0.33$. Since the scans shown in
Fig.~\ref{PNSXM_fig2} measure the amplitude of the magnetic
response it is possible that this broadening has washed-out the
$26$\,meV peak for $x=0.275$.

There is no sign of the dip--peak structure in the data for
$x=0.5$, and although the equivalent energy scans for $x=0$ do
present a superficial resemblance to those in
Fig.~\ref{PNSXM_fig2} the analysis in Ref.
\cite{Nakajima-JPSJ-1993} showed that for the case of $x=0$ the
signal in this energy range is fully accounted for by a spin wave
model with two anisotropy gaps. The anomalous dip--peak structure
in Figs.~\ref{PNSXM_fig2}(a)--(c) cannot be explained by two
anisotropy gaps because, as shown above, the anisotropy gaps for
$x=0.275$--0.37 are below 10\,meV.

\section{Discussion and conclusions}

\label{sec:disc} It is not clear at present what mechanism drives
the spin reorientation in La$_{2-x}$Sr$_{x}$NiO$_4$, but our
results show, (i) that the reorientation occurs for several doping
levels, including compositions in which the stripes are
commensurate ($x=0.33$) and incommensurate ($x=0.37$ and 0.5) with
the host lattice, and (ii) that the angle of the ordered moment
relative to the stripe direction tends to increase with doping.
The reorientation transition is most prominent for $x=0.5$, which
has the highest charge ordering temperature ($\sim 480$\,K) and a
slightly-incommensurate checkerboard charge ordering pattern below
$\sim 180$\,K \cite{Freeman-PRB-2002,Kajimoto-PRB-2003}.

We also cannot offer an explanation for the pronounced decrease in
single-ion anisotropy with doping. We can, however, comment on a
possible origin of the dip--peak feature in the spin excitation
spectrum. An explanation involving purely magnetic interactions
seems unlikely since the feature is most prominent for $x=0.33$,
whose spins order into a two-sublattice antiferromagnetic
structure \cite{Boothroyd-PRB-2003}. The zero-field spin wave
spectrum of such an ordering contains no gaps other than those due
to anisotropy, and we have shown above that for $x=0.33$ the
anisotropy gaps are 7\,meV and $\leq 1$\,meV. Moreover, the
polarization of the spin fluctuations is constant over the
dip--peak region, as illustrated in Fig.\ \ref{PNSXM_fig3}.
Hybridization with phonon excitations of the host lattice is a
possibility, but seems unlikely given that the anomalous dip is
absent for $x=0$ and $x=0.5$. Therefore, the evidence so far
points to a hybridization with an excitation associated with the
stripes, possibly a cooperative motion of the charge domain walls.

Finally, we mention that the experiments reported here illustrate
well the power of neutron polarization analysis to determine
static and fluctuating spin components, and to separate magnetic
and non-magnetic scattering unambiguously.

We thank F.R. Wondre for crystal characterization data, and the
EPSRC for financial support.


\begin{thebibliography}{00}
\bibitem{stripes-nickelates-expt}
C. H. Chen, S-W. Cheong, and A. S. Cooper, Phys. Rev. Lett. {\bf
71}, 2461 (1993); J. M. Tranquada {\it et al}, Phys. Rev. Lett.
{\bf 73}, 1003 (1994); H. Yoshizawa {\it et al}, Phys. Rev. B {\bf
61}, R854 (2000).

\bibitem{Tranquada-Nature-1995}
J. M. Tranquada {\it et al}, Nature {\bf 375} (1995) 561.

\bibitem{Boothroyd-PRB-2003}
A. T. Boothroyd {\it et al}, Phys. Rev. B {\bf 67} (2003)
100407(R).

\bibitem{Freeman-PRB-2002}
P. G. Freeman {\it et al}, Phys. Rev. B {\bf 66} (2002) 212405.

\bibitem{kulda-ApplPhysA-2002}
J. Kulda {\it et al}, Appl. Phys. A {\bf 74} [Suppl] (2002) S246.

\bibitem{Prabhakaran-JCG-2002}
D. Prabhakaran {\it et al}, J. Cryst. Growth {\bf 237} (2002) 815.

\bibitem{Lee-PRB-2001}
S.-H. Lee {\it et al}, Phys. Rev. B {\bf 63} (2001) 060405(R).

\bibitem{Nakajima-JPSJ-1993}
K. Nakajima {\it et al}, J. Phys. Soc. Jpn. {\bf 62} (1993) 4438.

\bibitem{Kajimoto-PRB-2003}
R. Kajimoto {\it et al}, Phys. Rev. B {\bf 67} (2003) 14511.






\end{thebibliography}
\end{document}